# BUBBLE SHAPE AND TRANSPORT DURING LCM PROCESSES: EXPERIMENTAL MODELING IN A T-JUNCTION TUBE

M. A. Ben Abdelwahed[1]*, Y. Wielhorski[1], L. Bizet[1], J. Bréard[1]

[1] Laboratoire Ondes et Milieux Complexes (LOMC), FRE 3102 CNRS, 53 rue de Prony, Université du Havre – BP 540, 76058 Le Havre cedex France

*Corresponding author (mohamed.ben-abdelwahed@univ-lehavre.fr)

**Keywords**: *Bubble, LCM process, capillary number, void formation, void transport.*

## 1 Introduction

Long fiber composite materials can be elaborated by Liquid Composite Molding (LCM), a family of processes where fibrous preforms are injected by a low viscosity resin. During this process, we have to pay particular attention to the void formation (Fig. 1a) inside the preform because it could modify the material final characteristics. Indeed, a preform presents two different porosity scales: between yarns called macropores and inside yarns, namely micropores. Due to this double porosity, bubbles of different shapes and lengths can be created inside composite materials [1]. It is observed experimentally that at low Capillary number *Ca*, the capillary flow is favored inside the yarns, i.e. between fibers, and leading to create inter tow voids or macrovoids. However, at higher *Ca*, the flow occurs between yarns is faster than the one between fibers at such velocities, the Stokes flow is more important and the associated regime included tow voids or microvoids [2]. Numerous numerical approaches [3, 4] attempted to simulate these voids creation by coupling Laplace forces and Stokes law.

However, the experimental visualization of the void formation and transport through the flow inside a fibrous preform remains delicate. Consequently, we have chosen to investigate the bubble generation and motion by a modeling device as a cylindrical capillary T-junction. It may represent for instance two convergent pores (Fig. 1b).

Many microfluidic flow-focusing devices are developed in order to study bubble creation. One of these devices is a rectangular T-shaped junction, which is used to create and characterize drops and bubbles by converging flows [5, 6]. Some studies attempt to build a flow phase pattern diagram linking the liquid capillary number and the gas flow rate [7]. Bubble velocities in capillary tube are also investigated [8, 9, 10, 11].

In the present work, we attempt by an experimental simple modeling to study bubble formation and transport mechanism for low Reynolds numbers ($Re \ll 1$).

## 2 Experimental procedures

In order to perform an advance-delay effect involved in bubble production phenomenon in LCM processes, we carry out an experiment which consists of converging two flows perpendicularly with different flow rates. Liquid is injected in the T-shaped junction by two syringe compressors in setting two different flow rates: $Q_1$, corresponding to the cross flow and $Q_2$, related to the gas injection (Fig. 2a). The break-up mechanism during the bubble formation is represented on Fig. 2b. The two flows merging at the junction create regular spaced bubbles. Bubble length *L* and distance between two successive bubbles *λ* are measured (Fig. 2c).

Glass capillaries are used to allow bubble visualization by a monochromic Dalsa M1024 camera. Accuracy of length measurement is about one pixel on recorded images. Thus precision of obtained values is around *40μm*. Images are analyzed with Aphelion 3.2 software. Three different liquids are used in our experiments: two silicone oils, Rhordorsil 47V100 and 47V1000 given by Rhodia with viscosities *η* of respectively 0.1Pa.s and 1.0Pa.s and a water-glycerol mixture in proportion (15-85%) with a viscosity of 0.1Pa.s. The liquid surface tensions $\gamma_L$ were measured by a K100 SF Krüss tensiometer for both silicone oils and the water-glycerol mixture. The values obtained are close to 21mN/m for the both silicone oils and 47mN/m for the mixture of water-glycerol. Two capillary tubes are used with two radii $R_c$ (0.5 and 1.0mm). This choice is governed by the capillary length with is close to 1.5mm for the silicone oils and about 2mm for the mixture water-glycerol.

# 3 Results and discussion

## 3.1 Bubble length

Fig. 3 shows the normalized bubble length in function of the flow rate ratio $Q_2/Q_1$. As the result, three different regimes [7] are distinguished according to the bubble length and the range of flow rate ratio values. Squeezing regime is defined for $L/2R_c>2.5$ in which long slug bubbles are obtained. In this regime, the interfacial force is much higher comparing to the cross-flow shear force and the dynamics of break-up is dominated by the filling pressure. Then, a transition regime where short slug bubbles are created, was observed for relative bubble length $1<L/2R_c<2.5$. Here, the break-up mechanism is dominated by a balance between both forces. Dripping regime, corresponding to the dispersed bubbles, is the third regime where relative bubble length $L/2R_c<1$.

Two different laws are found: for the squeezing and transition regimes (Eq. (1)) and for the dripping regime (Eq. (2)):

$$\frac{L}{2R_c} = \alpha \frac{Q_2}{Q_1} + \beta \quad (1)$$

$$\frac{L}{2R_c} = A \left(\frac{Q_2}{Q_1}\right)^B \quad (2)$$

with α, β, A and B are fitting coefficients.
Eq. (1) is a linear model where α and β depend on the liquid proprieties. However, for both silicone oils the Garstecki's model for square capillary microdevice [6] is very close to the experiment data. Eq. (2) is a power law relation in function of the flow rate ratio.

## 3.2 λ-model

To investigate the bubble frequency inside the T-junction device, which is strikingly related to both flow rates, distance between two successive bubbles noted λ is measured experimentally. This parameter could be quite important to understand bubble formation because it can determine the gas quantity created in function of time. In the following, a theoretical model named λ-model is proposed to determine the distance between two successive bubble centers by a geometrical approach.

Theoretical values are compared with experimental ones for both silicone oils and for $R_c=1$mm. The model is based on the determination of the liquid volume entrapped between two successive bubbles and the gas (bubble) volume, respectively noted $V_l$ and $V_b$. The elementary volume $V_\lambda$ corresponds to the whole volume delimited by two bubble centers noted by λ (Fig. 2c), i.e. the sum of $V_l$ and $V_b$. For transition and squeezing regime, the following equation is obtained:

$$\frac{\lambda}{2R_c} = \alpha \frac{Q_2}{Q_1} + \beta - \frac{1}{3} + \frac{K'}{2\pi R_c^3}\left(1+\frac{Q_1}{Q_2}\right) \quad (3)$$

Where $K'$ [m$^3$] is linked to the bubble time growth. Note that Eq. (3) is defined for $Q_2/Q_1 \geq 1.5$.

For dripping regime, the λ-model gives:

$$\frac{\lambda}{2R_c} = \frac{2}{3}\left(A\left(\frac{Q_2}{Q_1}\right)^B\right)^3 + \frac{K'}{2\pi R_c^3}\left(1+\frac{Q_1}{Q_2}\right) \quad (4)$$

Eq. (4) is valid for the condition $Q_2/Q_1 \leq 0.064$.

To conclude, we can say that the λ-model is in a good agreement with the experimental values (Fig. 4). Main result is that the distance between two successive bubble centers reaches a minimal value in the transition regime.

## 3.3 Bubble pattern

This part attempts to show the influences of different parameters on the squeezing-to-dripping diagram plotted with the cross flow capillary number $Ca_1$ and the gaseous phase flow rate $Q_2$ as coordinates. $Ca_1$ is defined as $\eta U_1/\gamma_L$ where $U_1$ is the main velocity for the cross flow $Q_1$. This study deals with regime limits according to the variations of bubble length. Regime limits was proposed by Fu and al. [7] but adapted to our cylindrical geometry.

The limit between transition and squeezing regimes ($Ca_1^{TS}$ for $L/2R_c = 2.5$) which is combined with Eq. (1) is given by:

$$Ca_1^{TS} = \frac{\alpha}{2.5-\beta}\frac{\eta}{\gamma_L}\frac{1}{\pi R_c^2}Q_2 \quad (5)$$

By the same reasoning, Eq. (2) is arranged to determine the limit between the dripping and transition regimes ($Ca_1^{TD}$ for $L/2R_c = 1$) is defined by the following capillary number:

$$Ca_1^{TD} = \left(\frac{1}{A}\right)^{-\frac{1}{B}} \frac{\eta}{\gamma_L} \frac{1}{\pi R_c^2} Q_2 \qquad (6)$$

Fig. 5 shows the capillary number $Ca_1$ as a function of the gas flow rate $Q_2$. The phase diagram obtained from the bubble length study is slightly different from the one obtained by Fu et al. [7] with different experimental parameters. Indeed, our experiments show that both gas and liquid flow rates influence bubble shape.

### 3.4 Relative bubble velocity

Fairbrother and Stubbs [8] showed that bubbles moved faster than the liquid. They think that it is probably due to the film thickness left by the liquid in movement inside the tube without measure it. They give an empirical relation (Eq. 7) linked the relative bubble velocity $(U_b-U_L)/U_b$ ($U_b$ is the bubble velocity and $U_L$ the main liquid velocity) and the capillary number which is defined as $Ca^* = \eta\, U_b/\gamma_L$ where $\eta$ the liquid dynamic viscosity and $\gamma_L$ the liquid surface tension:

$$\frac{U_b - U_l}{U_b} = 1.0\ (Ca^*)^{1/2} \qquad (7)$$

This model is valuable for the range $7.5 \cdot 10^{-5} < Ca^* < 0.014$.

Taylor [10] shows that for very viscous liquids moving at small velocities, the film thickness increases in function of the capillary number. He improved the Fairbrother and Stubbs' relation and found that bubble velocity not exceeds 2.27 times the liquid one at large $Ca^*$ and the film thickness cannot overtake $0.34 R_c$. Bretherton [9] quantified the movement of bubbles in circular capillaries by given a lubrication hypothesis. He found that the relative bubble velocity is given by:

$$\frac{U_b - U_l}{U_b} = 1.29\ (3Ca^*)^{2/3} \qquad (8)$$

Cox [12] solved numerically the problem concerning the fluid surrounding the bubble and presented experimental data indicating that the limit for large $Ca^*$ of the relative bubble velocity is 0.6. Thulasidas [11] found experimentally that this limit is about 0.58.

Landau, Levich and Derjaguin [13, 14] gave the lubrication hypothesis and showed that the film thickness is proportional to the capillary length ($\kappa^{-1}$) for a motion on a free surface and a function of $Ca^*$. They found a relation based on the continuity between the dynamic (*l*) and static meniscus which is weakly disrupted by the flow. This hypothesis links the film thickness and the junction length between the static meniscus and the deposition. Note that LLD hypothesis is validated when dynamic meniscus is only a little variation of static meniscus, i.e. when $l<<\kappa^{-1}$ or $Ca^*<<1$. Later, Marchessault and Mason [15] measured the film thickness surrounding the air bubbles inside circular capillaries by using a conductimetric technique. Schwartz et al. [16] studied the influence air bubble lengths on the liquid film surrounding them. Hitherto the most suggested models are given for low $Ca^*$ namely in the visco-capillary regime. Aussillous and Quéré [17] proposed a model for the film thickness including the inertial effect thanks to $We^*$ at higher $Ca^*$.

$$\frac{e}{R_c} \sim \frac{(Ca^*)^{2/3}}{1 + (Ca^*)^{2/3} - We^*} \qquad (9)$$

Where Weber number $We^*$ is defined by $2\rho U_b^2 R_c/\gamma_L$, with $\rho$ is the liquid density.

Fig. 6 shows the relative bubble velocity $(U_b-U_L)/U_b$ which is measured in a viscous flow and compared with semi-empirical models [8, 9] which based on the lubrication hypothesis. Our proposed model is function of $Ca^*$ and $We^*$.

$$\frac{U_b - U_l}{U_b} = \frac{a_1(Ca^*)^{2/3}}{1 + a_2(Ca^*)^{2/3} - kWe^*} \qquad (10)$$

The $We^*$ can be expressed as a function of $Ca^*$: $We^* = 2\gamma_L\rho R_c(Ca^*)^2/\eta^2$. Eq (10) so becomes:

$$\frac{U_b - U_l}{U_b} = \frac{a_1(Ca^*)^{2/3}}{1 + a_2(Ca^*)^{2/3} - a_3(Ca^*)^2} \qquad (11)$$

Where $a_1=2.27$, $a_2=3.35$ and $a_3=k^* 2\gamma_L\rho R_c/\eta^2=0.033$

The proposed relative bubble velocity model is valid for a range of $Ca^*$ from $10^{-4}$ to 1. As we can see on graphic at the upper left part of the Fig. 6, the Weber number $We^*$ begins to grow considerably from $Ca^*$ close to $10^{-2}$ and $10^{-1}$ respectively for silicone oils 47V100 and 47V1000 and about $6.10^{-3}$ for the water-glycerol mixture. Under these values of $Ca^*$, the visco-capillary regime is predominant and the term $(Ca^*)^2$ which is linked to the $We^*$ is weak. While above these values the inertial effects which are no more negligible have to be considered for the bubble velocities. The relative bubble velocity increases more rapidly from the consideration of the inertial effect.

### 3.5 Relative void fraction and influence of the bubble length on the pressure drop inside a channel

As the bubble lengths and frequencies have been studied, it can be deduced the ratio $L/\lambda$. It corresponds to the linear void fraction (Fig. 7) in comparison to an elementary volume $V_\lambda$.

The pressure drop per total flow rate depends on the relative void fraction.

$$\frac{\Delta P}{Q_1 + Q_2} = K\left(1 + \frac{L}{\lambda}\right) \quad (12)$$

with $K = 8\eta L/\pi R_c^4$

The Fig. 8 shows that the pressure drop depends on the bubble length. This means that the long slug bubble disrupts more the flow than the dispersed bubbles.

### 4 Conclusions

The squeezing-to-dripping transition during the bubble formation in a cylindrical T-junction device has been experimentally studied. It also observed that the bubble formation can be divided into three regimes: dripping, transition and squeezing regimes and these are correlated by the gaseous flow rate and the capillary number of the liquid phase. We defined these different kinds of regimes by quantifying the normalized bubble length which is defined by linear or power laws according to their shape. Then, we are attempted to determine the evolution of the distance between two successive bubbles by a λ-model which based on volume reasoning. The model is in a good agreement with the measurements. It shows that the values of λ are quite high during the dripping regimes where bubble frequency is low and decreases with the raise of $Q_2/Q_1$. A minimum is obtained during the transition regime and the distance increases again in the squeezing regime because the bubbles continue to grow with $Q_2/Q_1$. Considerations on film thickness led us to link it with the bubble velocity. Indeed, we observed that the bubble velocity is higher than the liquid one. Finally, another model representing the relative bubble velocity in function of $Ca^*$ is made from the lubrication hypothesis and inertial effect on the range of $Ca^*$ from $10^{-4}$ to 1. This model seems to show a convergence on the relative bubble velocity with the increasing of $Ca^*$. The aim for the next of the work is to link all the diagrams obtained for T-junction device with the issue of the void creation and transport inside LCM processes.

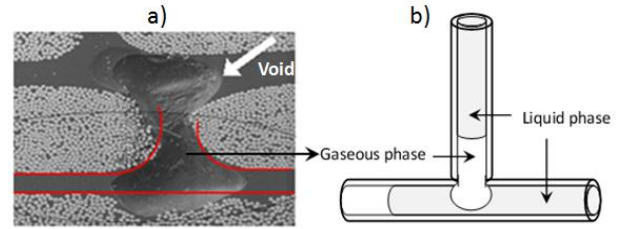

Fig. 1: a) Macro void in a fibrous preform, b) Experimental modeling of convergent macropores by a T-junction tube.

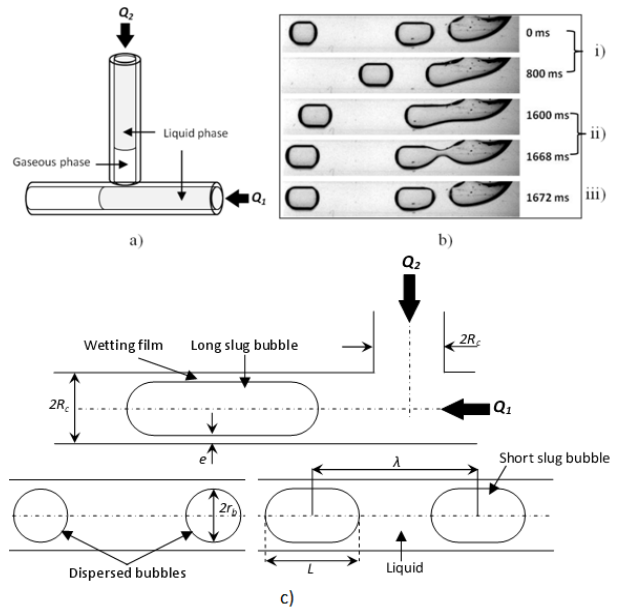

Fig. 2: (a) Advance-delay effect showing the gas entrapment; (b) Bubble formation: growth beginning (i), pinching (ii) and break-up (iii) mechanisms; (c) Bubble shapes in the different regimes.

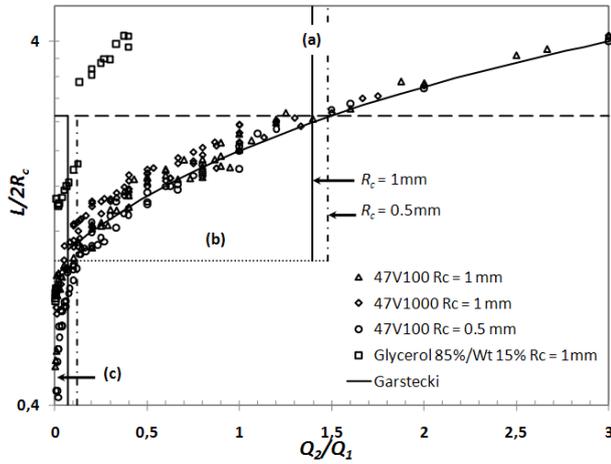

Fig. 3: Normalized bubble length as a function of flow rate ratio for: (a) squeezing regime; (b) transition regime; (c) dripping regime.

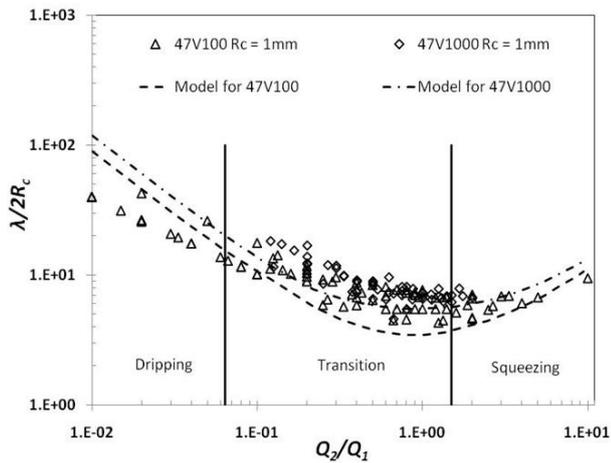

Fig. 4: Normalized distance between two successive bubbles as a function of flow rate ratio.

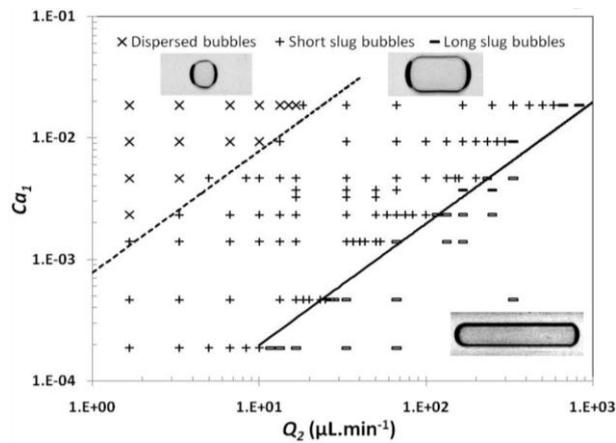

Fig. 5: Bubble shapes in the different regimes and their limits for 47V100 and $R_c$ = 1.0mm: dashed line is the transition line between dripping regime and transition regime; continuous line is the transition line between transition and squeezing regime.

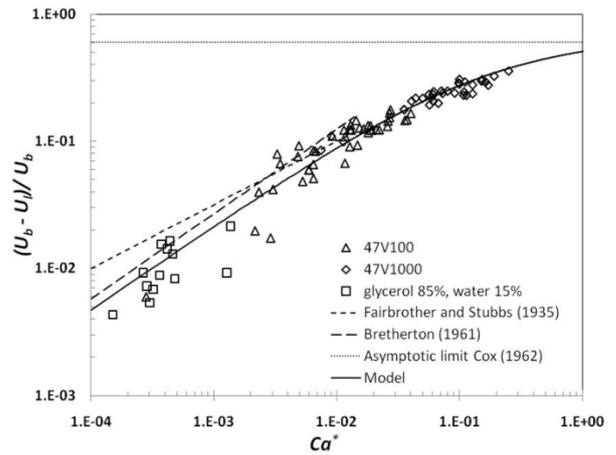

Fig. 6: Relative bubble velocity in function of capillary number $Ca^* = \eta U_b/\gamma_L$.

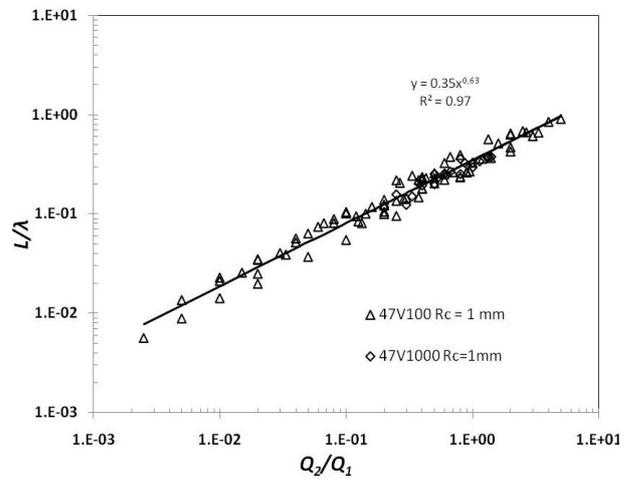

Fig. 7: Void fraction per distance between two bubbles in function of flow rate ratio.

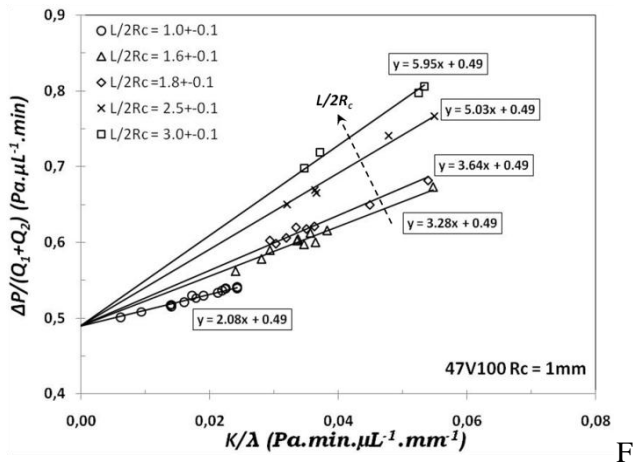

Fig. 8: Pressure drop per unit of total flow rate in function of the inverse distance between two successive bubbles.

## Acknowledgements

The authors gratefully acknowledge financial support from l'Agence Nationale de Recherche under the LCM3M project, the CNRS and la region de Haute Normandie.